\documentstyle[aps,12pt]{revtex}

\begin{document}
\author{A. C. Bertuola,$^{1}$ O. Bohigas$^{2}$ and M. P. Pato$^{1}$}
\address{$^{1}$Instituto de Fisica, Universidade de Sao Paulo\\
C.P.20516,01498 Sao Paulo, S.P., Brazil\\
$^{2}$Laboratoire de Physique Th\'{e}orique et Mod\`{e}les Statistique, \\
Universit\'{e} Paris-Sud, B\^{a}timent 100, \\
91405 Orsay Cedex, France.}
\title{Family of generalized random matrix ensembles }
\maketitle

\begin{abstract}
Using the Generalized Maximum Entropy Principle based on the nonextensive $q$
entropy a new family of random matrix ensembles is generated. This family
unifies previous extensions of Random Matrix Theory (RMT) and gives rise to
an orthogonal invariant stable L\'{e}vy ensemble with new statistical
properties. Some of them are analytically derived.
\end{abstract}

Random matrix theory (RMT) started in physics with the introduction by E.
Wigner, in the 50s, of Gaussian matrix ensembles, the Orthogonal (GOE), the
Unitary GUE) and the Symplectic (GSE). Their properties were fully developed
by Dyson, Gaudin, Mehta and others\cite{Meht}. These ensembles have a wide
application as models to describe statistical properties of quantum
fluctuations of systems of few or many-body particles. They have been useful
in discussing nuclear and atomic properties, mesoscopic physics, quantum
chaos, theory of amorphous solids, etc (see, for instance \cite{Guhr}). The
link between RMT and Information Theory was set by Balian\cite{Balian} who,
by using the Boltzman-Gibbs-Shannon entropy associated to the ensemble
probability distribution, obtained the Wigner ensembles by maximizing it
subjected to the normalization condition and a constraint given by the
average norm of the matrices. Ensembles to describe symmetry breaking have
been constructed by adding an extra constraint to this scheme\cite{Pato}.

In this letter, we use this framework and consider ensembles within the
Generalized Maximum Entropy Principle (GMEP) based on the nonextensive
Tsallis entropy\cite{Tsallis}. This entropy has been applied to a great
variety of phenomena, specially those in which long range correlations are
present (see however \cite{Nauen} concerning its physical interpretation).
It is dependent on the non-additivity parameter $q$ defined in such a way
that when $q\rightarrow 1$ the Boltzman-Gibbs-Shannon entropy is recovered.
We show that a new family of ensembles is generated that unifies some
important extensions of RMT. In the range $-\infty <$ $q<1$, it is found to
be a restricted trace ensemble that interpolates between the bounded trace
ensemble\cite{Bronk} when $q\rightarrow -\infty $ and the Wigner-Gaussian
ensembles at $q=1$. In the domain $1<q<q_{\max }$, with $q_{\max }$ being a
cutoff imposed by the normalization condition, it interpolates between RMT
at $q=1$ and an ensemble of L\'{e}vy matrices \cite{Cizeau} that appears at
the neighborhood of the extremum $q_{\max }$ where the ensemble distribution
has divergent moments.

As extensions of RMT that preserve the stability of the universal ensembles,
L\'{e}vy matrices have attracted recently much attention due to its
potential application to many areas ranging from physics to finances \cite
{Cizeau,Burda,Witte}. Stability means that if $H_{1}$ and $H_{2}$ are
matrices of the ensemble, their sum $H=H_{1}+H_{2}$ also is\cite{Tierz}.
This will be the case if the individual matrix elements are distributed
according to a Gaussian or a L\'{e}vy function. We prove that this indeed
happens, in the case of the $q$-generalized ensembles, for all allowed
values of $q$, i.e. $-\infty <q<q_{\max }$ when $N$ goes to infinity.

Although the individual matrix element distribution of $q$-ensembles have
the same asymptotic behavior as the L\'{e}vy matrices of Ref. \cite{Cizeau},
there is here a basic difference as they are orthogonal invariant with
matrix elements, in principle, correlated. Orthogonal invariance is also
satisfied by the ensembles of Refs. \cite{Burda,Witte} which are directly
defined in terms of the joint distribution of eigenvalues. However no
explicit reference to the matrix elements distribution is made there and the
spectral statistical measures are obtained expressing them in terms of
apropriately defined orthogonal polynomials. Here we do not apply this
technique and show that the special relation that $q$-ensembles have with
the Gaussian ensembles allows their spectral properties to be analytically
derived.

Applied to matrices whose entries are random variables, the nonextensive
entropy can be written as

\begin{equation}
S_{q}=\frac{1-\int dHP^{q}\left( H\right) }{q-1},  \label{1}
\end{equation}
where $H$ is a $N\times N$ matrix distributed according to $P\left( H\right) 
$ and $dH$ is the product of differentials of the independent variables of
the matrices. For definiteness we consider real symmetric matrices in which
case we have $f=N(N+1)/2$ independent matrix elements and the differential
in (\ref{1}) is conveniently defined as $dH=2^{N(N-1)/4}\prod_{1\leq i\leq
j\leq N}dH_{ij}$.

The GMEP consists in maximizing (\ref{1}) subjected to normalization

\begin{equation}
\int dHP\left( H\right) =1,  \label{4}
\end{equation}
and to the constraint\cite{Tsallis2} 
\begin{equation}
\int dHP^{q}\left( H\right) 
\mathop{\rm tr}%
H^{2}-\mu \int dHP^{q}\left( H\right) =0  \label{8}
\end{equation}
that fixes the $q$-average of the norm defined as the trace of the square of
the matrices. Following the usual steps of the variational method, we arrive
at the probability distribution

\begin{equation}
P(H;\lambda ,\alpha )=Z_{N}^{-1}\left( 1+\frac{\alpha }{\lambda }%
\mathop{\rm tr}%
H^{2}\right) ^{\frac{1}{1-q}}  \label{12}
\end{equation}
with $\lambda $ given by

\begin{equation}
\lambda =\frac{1}{q-1}-\alpha \mu =\frac{1}{q-1}-\frac{f}{2}.  \label{16}
\end{equation}
$Z_{N}$ in (\ref{12}) is the partition function and (\ref{8}) has been used
to determine the relation $\alpha =\frac{f}{2\mu }$ . Let us remark that had
we used Renyi's entropy\cite{Renyi} instead of Eq. (\ref{1}) we would also
have been led to Eq. (\ref{12}).

Changing from matrix elements to eigenvalue and eigenvector variables the
ensemble distribution factorizes and, after integrating over the eigenvector
parameters, we find for the eigenvalues the joint probability distribution

\begin{equation}
P(E_{1},...E_{N};\lambda ,\alpha )=K_{N}\left( 1+\frac{\alpha }{\lambda }%
\sum E_{k}^{2}\right) ^{\frac{1}{1-q}}\prod \left| E_{j}-E_{i}\right| ,
\label{26}
\end{equation}
where $K_{N}$ is the normalization constant. Taking $q\rightarrow 1$ in the
above, $\lambda \rightarrow \infty $ and the RMT distributions

\begin{equation}
P_{GOE}\left( H;\alpha \right) =Z_{GOE,N}^{-1}\exp \left( -\alpha 
\mathop{\rm tr}%
H^{2}\right)  \label{27}
\end{equation}
and

\begin{equation}
P_{GOE}(E_{1},...E_{N};\alpha )=K_{GOE,N}\exp \left( -\alpha \sum
E_{k}^{2}\right) \prod \left| E_{j}-E_{i}\right|  \label{29}
\end{equation}
are recovered.

Considering $-\infty <q<1,$ i.e. $-\frac{f}{2}>\lambda >-\infty $ the
condition $%
\mathop{\rm tr}%
(H^{2})=\sum E_{k}^{2}<-\frac{\lambda }{\alpha }$ has to be imposed in order
to warrant a real positive probability distribution for any $q$. These two
inequalities define hyperspheres in which the matrix elements and the
eigenvalues are confined in their respective spaces. Taking in Eq. (\ref{12}%
) the limit $q\rightarrow -\infty $ with the partition function given by

\begin{equation}
Z_{N}\left( q\right) =\left( -\frac{\pi \lambda }{\alpha }\right) ^{\frac{f}{%
2}}\frac{\Gamma \left( \frac{2-q}{1-q}\right) }{\Gamma \left( 1-\lambda
\right) }  \label{30}
\end{equation}
we find that the ensemble goes to the bounded trace ensemble 
\begin{equation}
P\left( H;-\frac{f}{2},\alpha \right) =\left( -\frac{\alpha }{\pi \lambda }%
\right) ^{\frac{f}{2}}\Gamma \left( \frac{f}{2}\right) \Theta \left( \frac{f%
}{2\alpha }-%
\mathop{\rm tr}%
H^{2}\right) ,  \label{34}
\end{equation}
where $\Theta \left( x\right) $ is the step function. The bounded trace
ensemble is known to follow the Wigner-Dyson statistics of the Gaussian
ensemble when $N\rightarrow \infty $. To show that this is also the case for 
$-\infty <q<1$ we consider the probability distribution of a generic matrix
element

\begin{equation}
p\left( x;\lambda ,\alpha \right) =\sqrt{-\frac{\alpha }{\pi \lambda }}\frac{%
\Gamma \left( 1-\lambda \right) }{\Gamma \left( \frac{1}{2}-\lambda \right) }%
\left( 1+\frac{\alpha }{\lambda }x^{2}\right) ^{-\frac{1}{2}-\lambda }
\label{62}
\end{equation}
and the correlation between two matrix elements $h_{1}$ and $h_{2}$

\begin{equation}
C\left( h_{1},h_{2}\right) =\left\langle h^{2}\right\rangle
^{2}-\left\langle \left( h_{1}h_{2}\right) ^{2}\right\rangle =\frac{1}{%
4\alpha ^{2}}\frac{\lambda ^{2}}{\left( 2-\lambda \right) \left( 1-\lambda
\right) ^{2}}.  \label{64}
\end{equation}
By taking the limit of large matrices, (\ref{62}) goes to the Gaussian
distribution

\begin{equation}
p\left( x;\lambda ,\alpha \right) \sim \sqrt{\frac{\alpha }{\pi }}\exp
\left( -\alpha x^{2}\right)  \label{66}
\end{equation}
while $C\left( h_{1},h_{2}\right) \rightarrow 0$ indicating that the matrix
elements behave as those of the Gaussian ensembles as $N\rightarrow \infty $%
. Numerical simulations\cite{Alberto} confirm that the level density is
given by the Wigner semi-circle law

\begin{equation}
\rho _{GOE}\left( E;\alpha \right) =\left\{ 
\begin{array}{c}
\frac{2\alpha }{\pi }\sqrt{\frac{N}{\alpha }-E^{2}},\qquad \left| E\right| <%
\sqrt{\frac{N}{\alpha }} \\ 
0,\qquad \left| E\right| >\sqrt{\frac{N}{\alpha }}
\end{array}
\right. .  \label{68}
\end{equation}
and spectral fluctuations follow GOE statistics.

Consider now $q>1$. The partition function is given by

\begin{equation}
Z_{N}\left( q\right) =\left( \frac{\pi \lambda }{\alpha }\right) ^{\frac{f}{2%
}}\frac{\Gamma \left( \lambda \right) }{\Gamma \left( \frac{1}{q-1}\right) }
\label{70}
\end{equation}
that requires the restriction $\lambda >0$ or $q<q_{\max }=1+\frac{2}{f}$.
We see that the introduction of the parameter $\lambda $ is crucial to be
able to study the limit $N\rightarrow \infty $. It maps the interval $%
1<q<q_{\max }$ onto the interval $\infty >\lambda >0.$ The Fourier transform
of the distribution of a generic matrix element, Eq. (\ref{62}), with $%
\lambda >0$ is

\begin{equation}
F\left( k;\lambda ,\alpha \right) =\sqrt{\frac{2}{\pi }}\frac{1}{\Gamma
\left( \lambda \right) }\left( k\sqrt{\frac{\lambda }{\alpha }}\right)
^{\lambda }K_{\lambda }\left( k\sqrt{\frac{\lambda }{\alpha }}\right)
\label{82}
\end{equation}
where $K_{\lambda }\left( z\right) $ is the modified Bessel function\cite
{Stegun}. In order to ensure that spectra scale independently of the size of
the matrices, $\alpha $ has to go to infinity when $N$ does. This can be
seen from the analytic expression of the level density, Eq. (\ref{130})
below. The requirement is that a characteristic value, say $E_{c}=\sqrt{%
\frac{N\lambda }{\alpha }}$, remains finite when $N$ diverges. In this
limit, $K_{\lambda }\left( z\right) $ can be replaced by its small $z$
expansion and keeping only the first terms we can write $F\left( k;\lambda
,\alpha \right) \sim \exp \left( -\Lambda \left| \frac{k}{2}\sqrt{\frac{%
\lambda }{\alpha }}\right| ^{\sigma }\right) $ with

\begin{equation}
\sigma =2\text{ and }\Lambda =\frac{1}{4\left( \lambda -1\right) }\text{%
\qquad if }\infty >\lambda >1.  \label{86}
\end{equation}
and

\begin{equation}
\sigma =2\lambda \text{ and }\Lambda =\frac{\Gamma \left( 1-\lambda \right) 
}{\Gamma \left( 1+\lambda \right) }\qquad \text{if }1>\lambda >0  \label{90}
\end{equation}
Therefore for $\infty >\lambda >1$ the distribution of a generic matrix
element approaches the Gaussian distribution 
\begin{equation}
p\left( x;\lambda ,\alpha \right) \simeq \sqrt{\frac{\left( \lambda
-1\right) \alpha }{\pi \lambda }}\exp \left[ -\left( \lambda -1\right) \frac{%
\alpha }{\lambda }x^{2}\right] .  \label{94}
\end{equation}
For $1>\lambda >0$ the L\'{e}vy-Gnedenko generalized central limit holds\cite
{Gnedenko} and $p\left( x;\lambda ,\alpha \right) $ goes to the L\'{e}vy
function, $L_{\sigma }\left( x,\sigma ,\Lambda \right) =\pi
^{-1}\int_{0}^{\infty }dt\exp \left( -\Lambda t^{\sigma }\right) \cos \left(
xt\right) $, with the same asymptotic behavior, i.e.

\begin{equation}
p\left( x;\lambda ,\alpha \right) \simeq 2\sqrt{\frac{\alpha }{\lambda }}%
L_{2\lambda }\left[ 2\sqrt{\frac{\alpha }{\lambda }}x,2\lambda ,\frac{\Gamma
\left( 1-\lambda \right) }{\Gamma \left( 1+\lambda \right) }\right] .
\label{98}
\end{equation}
Concerning correlations between matrix elements, Eq. (\ref{64}) shows that
only for large values of $\lambda $ or $\alpha $ the matrix elements behave
independently, whereas for small values, $\lambda <2,$ they are strongly
correlated. Therefore for large $\lambda $ or $\alpha $ (\ref{94}) predicts
that the level density goes to the semi-circle $\rho _{GOE}\left[ E;\left(
\lambda -1\right) \frac{\alpha }{\lambda }\right] $ i.e. Eq. (\ref{68}) with 
$\alpha $ replaced by $\left( \lambda -1\right) \frac{\alpha }{\lambda }$.

We focus now on the spectral properties of these new ensembles. They are
analytically derived by introducing the representation

\begin{equation}
\left[ 1+\frac{\alpha }{\lambda }%
\mathop{\rm tr}%
(H^{2})\right] ^{\frac{1}{1-q}}=\frac{1}{\Gamma \left( \frac{1}{q-1}\right) }%
\int_{0}^{\infty }d\xi \exp \left( -\xi \right) \xi ^{\frac{1}{q-1}-1}\exp
\left( -\frac{\alpha }{\lambda }\xi 
\mathop{\rm tr}%
H^{2}\right)  \label{102}
\end{equation}
that allows the joint distribution function of the matrix elements to be
written in terms of the joint distribution function of the GOE ensemble as

\begin{equation}
P\left( H;\lambda ,\alpha \right) =\frac{1}{\Gamma \left( \lambda \right) }%
\int_{0}^{\infty }d\xi \exp \left( -\xi \right) \xi ^{\lambda
-1}P_{GOE}\left( H;\frac{\alpha }{\lambda }\xi \right) ;  \label{106}
\end{equation}
the joint distribution of eigenvalues becomes

\begin{equation}
P\left( E_{1},...,E_{N};\lambda ,\alpha \right) =\frac{K_{N}}{\Gamma \left( 
\frac{1}{q-1}\right) }\int_{0}^{\infty }d\xi \exp \left( -\xi \right) \xi ^{%
\frac{1}{q-1}-1}\exp \left( -\frac{\alpha }{\lambda }\xi \sum
E_{k}^{2}\right) \prod \left| E_{j}-E_{i}\right|  \label{110}
\end{equation}
where $K_{N}$ is the normalization constant. Integrating (\ref{110}) over
all eigenvalues we deduce the relation

\begin{equation}
K_{N}=\frac{\left( \frac{2\alpha }{\lambda }\right) ^{\frac{f}{2}}\Gamma
\left( \frac{1}{q-1}\right) }{\Gamma \left( \lambda \right) }K_{GOE,N}
\label{118}
\end{equation}
relating $K_{N}$ to the corresponding RMT constant in standard units, i.e. $%
\alpha =\frac{1}{2}$ in Eq. (\ref{29}) see \cite{Meht}. Substituting in (\ref
{110}) one finally obtains for the normalized joint eigenvalue density

\begin{equation}
P\left( E_{1},...E_{N};\lambda ,\alpha \right) =\frac{1}{\Gamma \left(
\lambda \right) }\left( \frac{2\alpha }{\lambda }\right) ^{\frac{N}{2}%
}\int_{0}^{\infty }d\xi \exp \left( -\xi \right) \xi ^{\lambda +\frac{N}{2}%
-1}P_{GOE}\left( \sqrt{\xi }x_{1},...\sqrt{\xi }x_{N};\frac{1}{2}\right) 
\label{122}
\end{equation}
where we have introduced the rescaled eigenvalues $x_{k}=\sqrt{\frac{2\alpha 
}{\lambda }}E_{k}$. This is one of the central results of this paper and can
be taken as the defining equation of the new ensemble. It expresses the
eigenvalue distribution of the new ensemble as a sort of $\Gamma $ function
of the GOE eigenvalue distribution. It shows that one may expect that
measures of the $q$-family will be weighted Laplace transforms of the
corresponding measures of the Gaussian ensemble.

Integrating (\ref{122}) over all eigenvalues but one and multiplying by $N,$
the average eigenvalue density is expressed in terms of Wigner's semi-circle
law as

\begin{equation}
\rho \left( E;\lambda ,\alpha \right) =\frac{1}{\Gamma \left( \lambda
\right) }\sqrt{\frac{2\alpha }{\lambda }}\int_{0}^{\frac{N\lambda }{\alpha
E^{2}}}d\xi \exp \left( -\xi \right) \xi ^{\lambda -\frac{1}{2}}\frac{1}{\pi 
}\sqrt{2N-2\frac{\alpha }{\lambda }\xi E^{2}}.  \label{126}
\end{equation}
The asymptotic power law behavior of this distribution is better seen by
rewriting it as

\begin{equation}
\rho \left( E;\lambda ,\alpha \right) =\frac{N}{\left| E\right| ^{2\lambda
+1}\sqrt{\pi }}\left( \frac{N\lambda }{\alpha }\right) ^{\lambda }\frac{%
\Gamma \left( \lambda +\frac{1}{2}\right) }{\Gamma \left( \lambda \right)
\Gamma \left( \lambda +2\right) }M\left( \lambda +\frac{1}{2},\lambda +2,-%
\frac{N\lambda }{\alpha E^{2}}\right)  \label{130}
\end{equation}
where $M\left( a,b,z\right) $ is the confluent hypergeometric function\cite
{Stegun}. In Fig. 1, with $\alpha =\frac{N^{\frac{2}{\sigma }}}{2}$ (see
Eqs. (\ref{86}) and (\ref{90})) the density $\rho \left( E;\lambda ,\alpha
\right) $ is plotted for four values of $\lambda $, exhibiting the deviation
from the semi-circle law as $\lambda $ moves inside the interval $1>\lambda
>0$. When $\lambda \rightarrow 0$, the density behaves as $\rho \simeq $ $%
\frac{N^{\lambda }}{\left| E\right| ^{2\lambda +1}}$ approaching the same
behavior as for a nonconfining log square potential\cite{Bogom}.

The behavior of the spectral fluctuations can be illustrated by considering
the gap probability function $E\left( s\right) $ (usually denoted $E\left(
0,s\right) $) that gives the probability of finding an eigenvalue-free
segment of length $s.$ This function has been investigated in Ref. \cite
{Witte} for Cauchy ensembles and is related to the presence of gaps in the
spectrum. For the $q$-family it is expressed in terms of the corresponding
GOE function as

\begin{equation}
E\left( \theta \right) =\frac{1}{\Gamma \left( \lambda \right) }%
\int_{0}^{\infty }d\xi \exp \left( -\xi \right) \xi ^{\lambda
-1}E_{GOE}\left[ y\left( \sqrt{\frac{2\alpha \xi }{\lambda }}\theta \right)
\right]  \label{146}
\end{equation}
obtained integrating the joint eigenvalue density over all eigenvalues
outside the interval $\left( -\theta ,\theta \right) $ around the origin. In
(\ref{146}) $y\left( x\right) =2\int_{0}^{x}dt\rho _{GOE}\left( t\right) .$
Together with

\begin{equation}
s\left( \theta \right) =2\int_{0}^{\theta }dE\rho \left( E;\lambda ,\alpha
\right)  \label{148}
\end{equation}
(\ref{146}) expresses $E\left( s\right) $ in a parametric form. Using the
Wigner surmise for the nearest neighbor spacing distribution $p\left(
s\right) $ and the relation connecting $E\left( s\right) $ and $p\left(
s\right) $, $E_{GOE}$ in (\ref{146}) can be well approximated by $%
E_{GOE}\left( y\right) \simeq 1-%
\mathop{\rm erf}%
\left( \frac{y\sqrt{\pi }}{2}\right) $. On Fig. 2 results in the L\'{e}vy
regime are displayed. Notice the large increase of the probability of
formation of a gap with respect to the GOE case. The asymptotic behavior in
Eq. (\ref{146}) can be extracted by making the substitution $x=\sqrt{\frac{%
2\alpha \xi }{\lambda }}\theta $ that leads to

\begin{equation}
E\left( \theta \right) =\frac{2}{\Gamma \left( \lambda \right) }\left( \frac{%
\lambda }{2\alpha }\right) ^{\lambda }\frac{1}{\theta ^{2\lambda }}%
\int_{0}^{\infty }dx\exp \left[ -\frac{\lambda }{2\alpha }\left( \frac{x}{%
\theta }\right) ^{2}\right] x^{2\lambda -1}E_{GOE}\left[ y\left( x\right)
\right] .  \label{150}
\end{equation}
For large $\theta ,$ this equation predicts for $\lambda =1$ a power law
decay $E\left( s\right) \simeq \frac{1}{2s^{2}}$, clearly seen in the
figure. This very characteristic behavior is exhibitted here for the first
time.

In summary, we have proved that the $q$-generalized family of ensembles
interpolates between the bounded trace ensemble\cite{Bronk} at the extremum $%
q\rightarrow -\infty $ and the Wigner-Gaussian ensembles at $q=1.$ In the
domain $1<q<q_{\max }$, it interpolates between RMT at $q=1$ and an ensemble
of L\'{e}vy matrices at the neighborhood of the extremum $q_{\max }=1+\frac{2%
}{f}$. These orthogonal invariant stable matrix ensembles have novel
spectral properties. Remarkably, several of their distribution functions can
be expressed as integral transforms (sort of extended $\Gamma $ functions)
of the corresponding distribution functions of the Gaussian ensembles.

It is premature to exhibit specific applications of these generalized
ensembles. However there are worth exploring possibilities, for instance,
connections with the so-called critical statistics\cite{Garcia} or the
transition from Erd\"{o}s-Renyi to scale free models in random graph theory%
\cite{Barab}. In conclusion, let us remind that stable laws (L\'{e}vy laws)
were first introduced and studied. It was correctly anticipated\cite
{Gnedenko} that a large domain of applications would follow\cite{Levy}. We
believe that we are presently facing a similar situation, where the role of
a random variable is now being extended to the one of a random matrix. The
results presented here should contribute to broaden the applications of
random matrix theory.

After completion of this letter, we learned of ref. \cite{Toscano} closely
related to the work presented here.

Fruitful discussions with C. Tsallis are acknowledged. A.C.B. and M.P.P. are
supported by the Conselho Nacional de Pesquisas (CNPq). This work is
supported by a project CAPES-COFECUB.

{\bf Figure Captions}

Fig. 1 The eigenvalue density for four values of the parameter $\lambda $ ($%
=10$, $1$, $0.75$, $0.5$) in the transition region from the Gaussian to the
L\'{e}vy regime, with $N=50$. For the sake of comparison the semi-circle $%
\rho _{GOE}\left[ E;\left( \lambda -1\right) \frac{\alpha }{\lambda }\right] 
$ with $\lambda =10$ and $\alpha =\frac{N^{\frac{2}{\sigma }}}{2}$ is also
shown (dashed line).

Fig. 2 The eigenvalue-free probability $E\left( s\right) $ for $\lambda =1$.
Full line: theory, Eq. (\ref{146}) and its asymptotics (dotted line); dashed
line: $E_{GOE}\left( s\right) $; * : numerical simulation with $N=20$. See
text for further explanation.

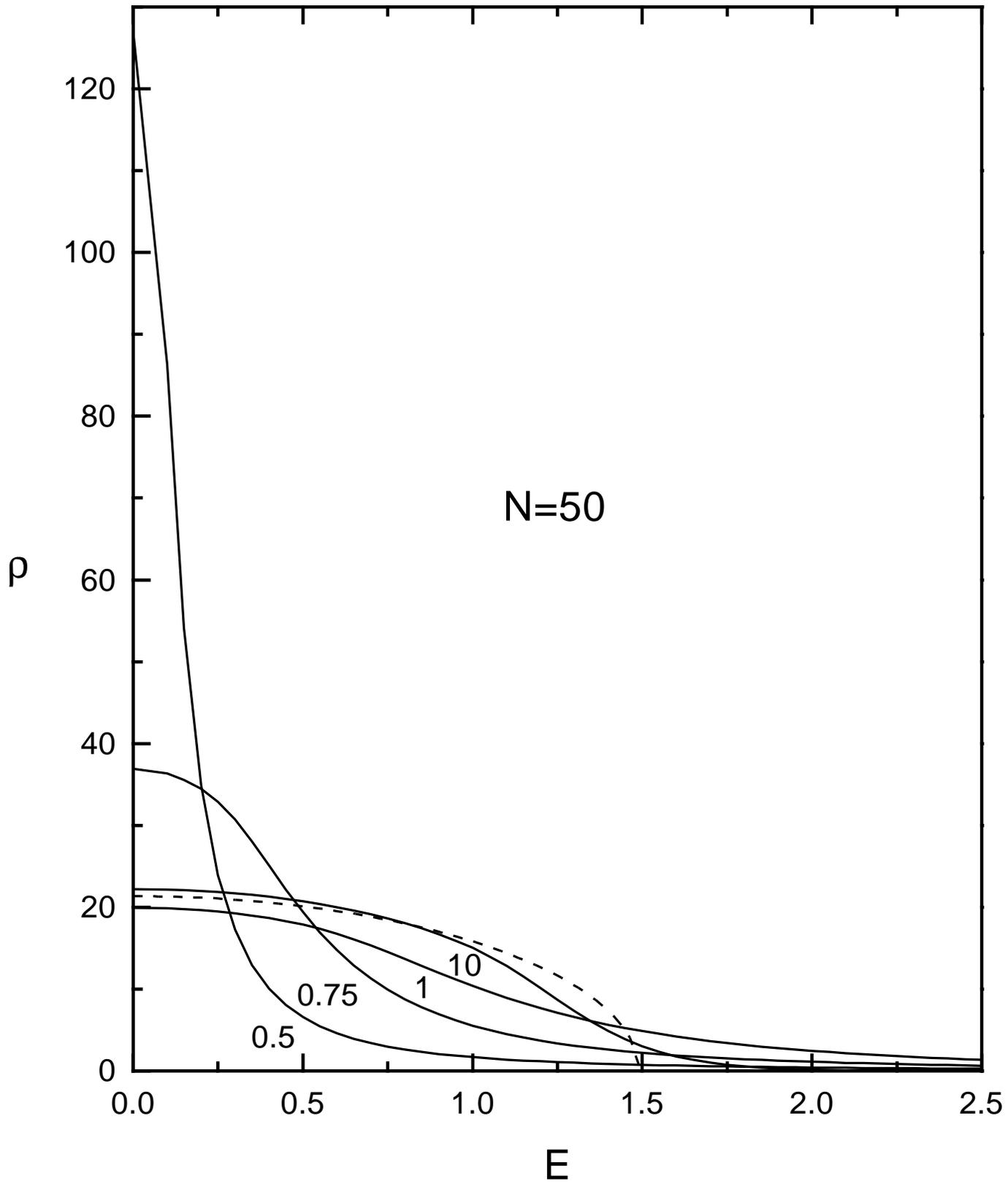
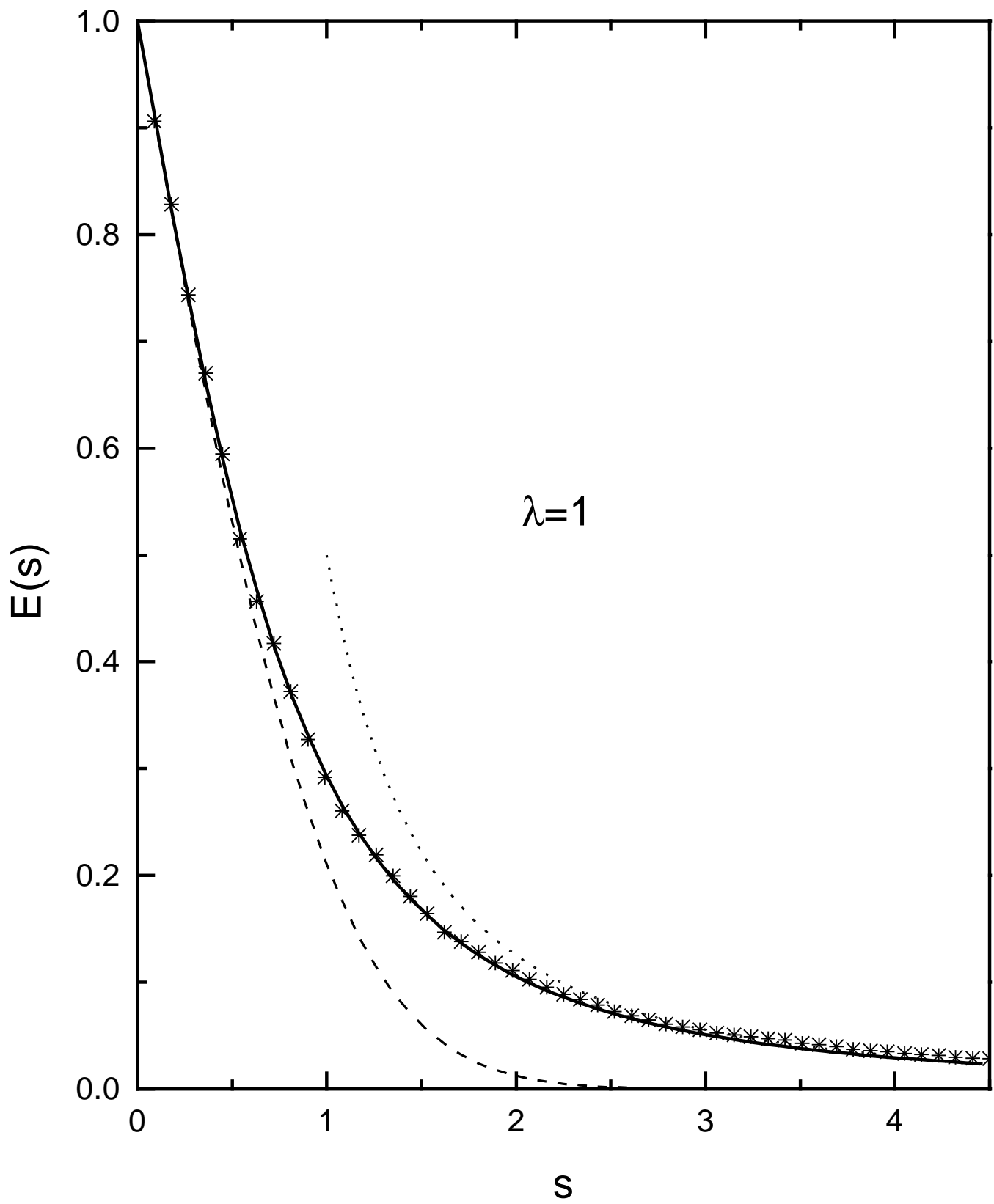
\end{document}